\documentclass[preprint,12pt]{elsarticle}

\usepackage{amssymb}
\usepackage{amsthm}

\usepackage{amsmath}
\usepackage{subcaption}

\journal{ }

\begin{document}

\begin{frontmatter}



\title{Advanced Scaling Methods for VNF deployment with Reinforcement Learning}

\author[hgu]{Namjin Seo}
\author[hgu]{DongNyeong Heo}
\author[hgu]{Heeyoul Choi}

\affiliation[hgu]{organization={Handong Global University},
            city={Pohang},
            postcode={37554},
            state={Gyeongbuk},
            country={South Korea}}






\begin{abstract} 
Network function virtualization (NFV) and software-defined network (SDN) have become emerging network paradigms, allowing virtualized network function (VNF) deployment at a low cost. Even though VNF deployment can be flexible, it is still challenging to optimize VNF deployment due to its high complexity. 
Several studies have approached the task as dynamic programming, e.g., integer linear programming (ILP). However, optimizing VNF deployment for highly complex networks remains a challenge.
Alternatively, reinforcement learning (RL) based approaches have been proposed to optimize this task, especially to employ a scaling action-based method which can deploy VNFs within less computational time. However, the model architecture can be improved further to generalize to the different networking settings. 
In this paper, we propose an enhanced model which can be adapted to more general network settings. We adopt the improved GNN architecture and a few techniques to obtain a better node representation for the VNF deployment task. Furthermore, we apply a recently proposed RL method, phasic policy gradient (PPG), to leverage the shared representation of the service function chain (SFC) generation model from the value function. 
We evaluate the proposed method in various scenarios, achieving a better QoS with minimum resource utilization compared to the previous methods. 
Finally, as a qualitative evaluation, we analyze our proposed encoder's representation for the nodes, which shows a more disentangled representation.  
\end{abstract}



\begin{keyword}
Network Function Virtualization \sep Software Defined Networking \sep Reinforcement Learning \sep Graph Neural Network
\end{keyword}

\end{frontmatter}


\section{Introduction}

Softwarization of the Internet network, such as software-defined networking (SDN) and network function virtualization (NFV), has emerged as a new network paradigm. In this paradigm, the network functions provided by hardware-based middleboxes (e.g., Network Address Translation (NAT), Firewall, Proxy) are replaced with virtualized network functions (VNFs), running on virtual machines as VNF instances. By decoupling the network functions from the hardware, NFV allows the network service providers to deploy VNFs with low capital expenses (CAPEX) and operating expenses (OPEX). In addition, the traffics and network devices are managed and monitored by NFV orchestration (NFVO) in the centralized NFV framework. Hence, the VNF instances can be deployed by NFVO adaptable to the traffic requirements, which is the VNF deployment task.

\begin{figure}[t]
    \centering
    \includegraphics[width=1\textwidth]{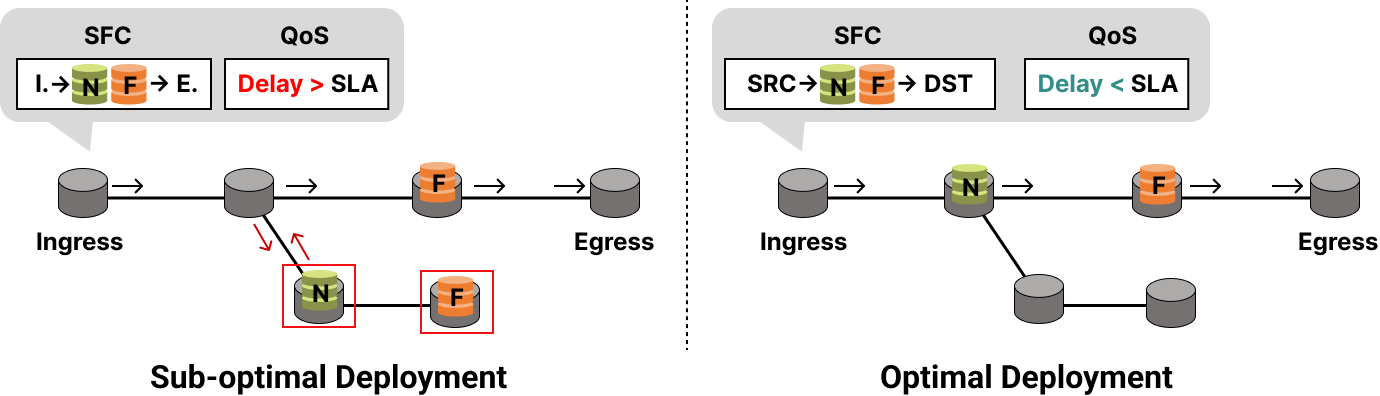}
    \caption{Overview of VNF Deployment task: SFC requests are required to pass through the VNFs sequentially as well as satisfy the SLA. Optimal deployment (Right) takes into account the paths of SFC requests, while sub-optimal deployment (Left) is deployed regardless of the paths of SFC request. Red boxes indicate inefficiently deployed VNF instances, which are on the wrong path and redundant.}
    \label{fig:sfc}
\end{figure}

Even though VNF deployment can be accomplished dynamically with flexibility, requirements for the VNF deployment task are also getting more complex. The first requirement is that the service function chain (SFC) should be generated efficiently while maintaining an acceptable quality of service (QoS). SFC requires the traffics of a request to be routed through multiple stages of VNFs in NFV. 
Fig. \ref{fig:sfc} shows an example of SFC where the SFC request ($\text{Req}_1$) demands an SFC: ``NAT (`N') $\rightarrow$ Firewall(`F')''.
This SFC request should be sequentially processed at these VNF instances as it travels from its ingress to its egress. These types of VNF should be deployed, taking into account the path of the requests, to meet the pre-defined service-level agreement (SLA) for QoS. Another requirement is to optimize resource utilization while satisfying the QoS. Redundantly deployed VNF instances could preserve high QoS, but it incurs unnecessary operating costs and energy consumption. Therefore, it is essential to improve the QoS while maintaining minimized resource consumption for the efficient management of VNF deployment.  

To achieve such efficient management of VNF deployment, the existing works have exploited dynamic programming algorithms, like integer linear programming (ILP) \cite{bari2015ilp,  moens2014vnfp, abdel2021il}. However, even though the ILP-based approach exhibits acceptable performance in networks with a low level of complexity, its computational cost becomes too expensive as the network scales up. Therefore, the ILP-based approach is not suitable for large-scale networks and the dynamical adjustment for traffic lifespans. 

As an alternative solution, ML-based approaches have been proposed employing deep learning models, and reinforcement learning (RL) for VNF deployment \cite{lange2017multi, farshi2020servplace, fernando2021addpg, sun2021gnn}. 
Especially, an RL-based method for VNF deployment was proposed to scale in and out VNFs on the network nodes \cite{namjin2022ppodepl}. 
They adapted the graph neural network (GNN) model and RL algorithms on top of previously deployed VNF instances. Their action controls the number of VNFs on each node by scaling in, scaling out, or doing nothing. It showed that the agent was able to make appropriate scaling decisions for all the nodes and VNF types with a single forwarding process, involving fewer computation times for VNF deployment compared to other ML-based VNF deployment approaches. However, their model could not obtain the node representations in a general deployment setting where the networks include some nodes which cannot have VNF instances (e.g., switches) deployed on them due to its model setting for the node features. 



In this paper, we propose to enhance the neural network model and adapt new learning techniques so that the model can be adapted to the general deployment setting. 
First, we redesign the GNN architecture, and adapt a few techniques motivated from other domains, such as natural language processing and image processing \cite{cho2014encdec, bahna2014attn, vaswani2017transf, alex2020vit, Lee:partition_repre}, to effectively obtain a node representation of the networking information for the VNF deployment task.
As the new GNN-based architecture, we employ the graph attention network (GAT) \cite{velickovic2018gat} and separately process different types of nodes: VNF deployable or non-deployable nodes. This architecture allows an effective propagation of node information from the network with different types of nodes. In addition, we adapt the positional encoding \cite{vaswani2017transf} to preserve the positional information in the node representations. 

Furthermore, we apply a recently proposed RL algorithm, phasic policy gradient (PPG) \cite{cobbe2020ppg}, which is a variation of the policy gradient algorithm to optimize the policy network efficiently. 
By jointly optimizing the objectives of the policy network and the value network, the policy network can share representation with the value network, which helps the policy network obtain a more effective representation for the VNF deployment task.


In the experiment, our approach optimizes the policy network to obtain a higher reward with rapid convergence compared to the previous approaches. Also, it proves that our approach can work in various scenarios. In addition, we analyze the result of our approach to show the robustness in the various topology and compare the processing time with ILP to show the practicality for the dynamic adaptation of the traffic lifespans. Finally, as a qualitative evaluation, we analyze the representation from the encoder of the policy network, and it demonstrates that our method obtains more disentangled representations of nodes.


\section{Background}
In this section, we briefly review graph neural networks and reinforcement learning for VNF deployment. 

\subsection{Graph Neural Network}
Graph neural networks (GNN) \cite{scarselli:gnn} were proposed to effectively handle the graph-structured data consisting of a set of nodes $\mathcal V = \{v_1, \cdots, v_{|\mathcal V|}\}$ and edges $\mathcal E = \{\hat e_{ij} | v_i, v_j \in \mathcal V \}$ \cite{zhou2021gnnreview, kipf2016gcn, li2015ggnn, velickovic2018gat, gilmer2017mpnn, wang2019heterogat}. GNNs aim to address graph-related tasks (e.g., node/edge/graph classification) in an end-to-end manner with neural network model \cite{zhou2021gnnreview}. In the training process, GNNs are trained to extract nodes' representations which reflect the graph information for the target tasks. 

In GNNs, a graph is represented as the following two matrix forms: the adjacency matrix $\mathbf{A}$ and the node feature matrix $\mathbf{X}$. The adjacency matrix $\mathbf{A} \in \mathbb{R}^{|\mathcal V| \times |\mathcal V|}$ represents the connections between pairs of nodes. The node feature matrix $\mathbf{X}$ contains node features $\mathbf{x}_i \in \mathbb{R}^{|\mathcal V| \times d }$. 
With $\mathbf{A}$, $\mathbf{X}$ is transformed into node representation $\mathbf{H}=[\mathbf{h}_1, ... ,\mathbf{h}_{|\mathcal{V}|}]$. Finally, these node representations are adapted to the target task in an end-to-end manner, which is implemented with a multi-layer perceptron (MLP) and a softmax layer. 

GNN models are designed under the assumption that a node representation for the target task is trained by propagating node information to its neighbors \cite{zhou2021gnnreview}. There are many different types of GNN models depending on the way how the node information propagates, like recurrent connection, convolution, or attention.  
Among many GNN models, we review and compare gated graph neural network (GGNN) and graph attention network (GAT) which are used in the baseline and our approach, respectively. 

\subsubsection{Gated Graph Neural Network}
In gated graph neural network (GGNN) \cite{li2015ggnn}, propagation between neighbors is implemented as recurrent connection as in recurrent neural networks (RNNs), and the node representation is calculated iteratively by RNN-like updates. For the recurrent connection, they use gated recurrent unit (GRU) \cite{cho2014encdec}, and the node representations are updated as follows: 
\begin{equation}
    \mathbf{h}'_i = GRU(\mathbf{h}_i, \sum_{j\in \mathcal N_i} \mathbf{W}\mathbf{h}_j),
\end{equation}
where $GRU$ is the GRU-like update function, $\mathcal N_i$ are the neighbors of node $v_i$, 
$\mathbf{W}$ is a weight matrix, and the initial node representation is set to $\mathbf{X}$. The recurrent updates are repeated a fixed number of times. Then, node representations are fed into the output layers.   

\subsubsection{Graph Attention Network}
Graph attention network (GAT) \cite{velickovic2018gat} was proposed to update node representations with attention mechanism \cite{bahna2014attn}, while node information is propagated through recurrent steps in GGNN. GAT computes the attention scores of its neighbors so that the node representations can be updated according to the different importance of its neighbors as follows:
\begin{align}
    \mathbf{h}'_i &= \sum_{j\in \mathcal N_i \cup \{i\}} \hat\alpha_{ij} \mathbf{W}\mathbf{h}_i, \label{eqn:gat1} \\
    \hat \alpha_{ij} &= \frac{\exp(\sigma(\mathbf{a}^\top\left[ \mathbf{W}\mathbf{h}_i||\mathbf{W}\mathbf{h}_j||\mathbf{W}_e \mathbf{\hat e}_{ij} \right] ))}
            {\sum_{j'\in \mathcal N_i \cup \{i\}} \exp(\sigma(\mathbf{a}^\top\left[ \mathbf{W}\mathbf{h}_i||\mathbf{W}\mathbf{h}_{j'}||\mathbf{W}_e\mathbf{\hat e}_{ij'} \right] ))}, \label{eqn:gat2}
\end{align}
where $\alpha_{ij}$ is the attention score of node $v_j$ for node $v_i$, and $\mathbf{ \hat e}_{ij}$ is edge attributes between nodes $v_i$ and $v_j$.  $\mathbf{W},\mathbf{W}_e$, and $\mathbf{a}$ are weight matrices, $\sigma$ is a non-linear function, and $\left[ \cdot || \cdot\right]$ indicates concatenation. 


\subsection{Reinforcement Learning}

Reinforcement learning (RL) trains an agent to maximize the expected reward by interacting with the environment. It formulates the task as Markov decision process (MDP) described as a sequence of state $s_t$, action $a_t$, and reward $r_t$ over discrete time steps. 
At discrete time-step $t$, the agent observes the state $s_t$ of the environment and takes an action $a_t$. Subsequently, the current state $s_t$ transitions into the next state $s_{t+1}$ along with the transition probability $P(s_{t+1}| s_t, a_t)$, and finally, the agent receives a reward $r_{t+1}$. This process is repeated until the agent reaches the terminal state which ends one episode. 
As the agent experiences many episodes, the agent could maximize the return, $G_t = r_{t+1} + \gamma r_{t+2} + \gamma^2r_{t+3} + \cdots$, where $\gamma$ is the discount factor. 

To maximize $G_t$ over an episode, the RL agent needs to update policy $\pi(a_t|s_t)$, a probability distribution of possible actions given state $s_t$. To estimate how good the action is at discrete time $t$, the agent should learn the state value $V(s_t)=\mathbb{E}[G_t|s_t]$ (or action-state value $Q(s_t,a_t)$ at action-state pair $(a_t,s_t)$).
In other words, the goal of an RL agent is to optimize its $\pi(a|s)$ based on the trained state value $V(s)$ (or action value $Q(s,a)$) to maximize the expected return in any episode. 

\subsubsection{Policy Gradient}

Policy gradient (PG) is an RL method to model policy $\pi(a|s)$ as a parameterized function with neural networks called the policy network $\pi_{\theta}(a|s)$. It trains the parameter of the policy network optimizing the objective function defined as follows:
\begin{equation}\label{eqn:obj_pg}
\mathcal L^{PG} (\theta) = \mathbb E_t \left[ \log \pi_\theta(a_t|s_t) \hat A_t \right],
\end{equation}
where $\hat A_t$ is the advantage estimator. For example, the REINFORCE algorithm \cite{sutton1999pg} trains the parameters of the policy network defining advantage estimator $\hat A_t$ as return $G_t$. 

\subsubsection{Proximal Policy Optimization}

Proximal policy optimization (PPO) \cite{shulm2017ppo} is one of the actor-critic (AC) based RL algorithms \cite{konda1999ac} with the trust-region method \cite{conn2000trustregion}. In addition to the policy network, the AC algorithm employs the value network $V_{\theta}(s)$ estimating the state value $V(s_t)$ in order to mitigate the variance in the training process. 
Furthermore, PPO has an additional constraint on the step size of the policy update to prevent inappropriate update of parameters, called the trust-region method \cite{conn2000trustregion, shulm2015trpo}. Let $\hat{r}_t(\theta)$ be the probability ratio $\hat{r}_t(\theta) = \frac{\pi_{\theta}(a_t|s_t)} {\pi_{\theta_{old}}(a_t|s_t)}$, then PPO maximizes
\begin{equation}\label{eqn:obj_ppo}
    \mathcal L^{PPO}(\theta) = \mathbb E_t \left[
    \text{min}(\hat{r}_t(\theta) \hat{A}_t, 
    \text{clip}(\hat{r}_t(\theta), 1-\epsilon, 1+\epsilon)\hat{A}_t) \right],
\end{equation}
where clip$(\cdot, x, y)$ is an operation clamping in range over $x$ and $y$, and $\epsilon$ is a hyper-parameter of the clipping. The clipping operation constrains the probability ratio $\hat{r}_t$ to be close to 1 so that the parameters cannot be updated radically.


\subsubsection{Phasic Policy Gradient}

The phasic policy gradient (PPG) \cite{cobbe2020ppg} algorithm was proposed to share parameters between the policy and value networks in the AC algorithms. 
Even though parameter sharing allows both networks to learn features jointly with a higher level of sample reuses, it is unclear whether it optimizes efficiently both networks jointly or not. Thus, they employ two value networks to avoid conflicts between competing objectives of the policy and value networks. The one is value network $V_{\theta_V}(s)$ not sharing parameters, and the other shares the parameter, called the auxiliary value head $V_{\theta_{\pi}}(s)$.

The PPG algorithm divides the policy and value network training phases into policy and auxiliary phases. In the policy phase, the policy is optimized in the same manner as PPO. Then, in every ${N_{PPG}}^{th}$ iterations, an auxiliary loss is minimized in the auxiliary phase, and the auxiliary loss is defined by
\begin{equation}\label{eqn:aux_loss}
    \mathcal L^{aux} = \mathbb{E}_t\left[ \frac{1}{2}(V_{\theta_{\pi}}(s_t)- V_{\theta_V}(s_t))^2\right] + \beta_{clone} \mathbb{E}_t\left[ KL\left[ \pi_{\theta_{old}}(\cdot|s_t),\pi_{\theta}(\cdot|s_t)  \right]\right],
\end{equation}
where $\beta_{clone}$ is the hyper-parameter controlling how much the original policy is preserved.

\subsection{Scaling Agent for VNF Deployment} \label{sec:scaling}


Scaling of VNF deployment monitors VNF instances deployed in a network and adjusts the number of instances to maintain reliable QoS at minimized cost by scaling in (removing instances), keeping, and scaling out (adding new instances).
An RL-based method was proposed for VNF scaling \cite{namjin2022ppodepl} to optimize QoS and resource utilization, simultaneously. 
They adapted the graph neural network (GNN) models and RL algorithms on top of previously deployed VNF instances. The GNN-based model inputs the network information, and then produces the probabilities of scaling actions: scale in/out and keep. The parameters of the model were updated by PG-based RL algorithms, such as REINFORCE \cite{sutton1999pg} and PPO \cite{shulm2017ppo}. In this section, we briefly review the model of the policy network and the RL formulation in \cite{namjin2022ppodepl}.


\subsubsection{GNN based Policy Network}

The policy network has an encoder-decoder architecture. The encoder receives the network information (e.g., network topology, deployed VNFs, and SFC requests.) and outputs the node representation reflecting the network information. The decoder reads the representations and produces action probabilities for `scaling in/out and keep' of nodes and VNF types. To compute node representations that reflect diverse network topologies, the encoder is designed based on GNN \cite{SFCmodel2020Heo}, which is implemented with Gated Graph Neural Network (GGNN) \cite{li2015ggnn}. Additionally, the network topology-related knowledge is transferred from the trained model for the SFC task, and the parameters are frozen so that the model could obtain proper representations of the network information.

\subsubsection{RL formulation for VNF scaling}

\begin{figure}[t]
    \centering
    \includegraphics[width=0.9\textwidth]{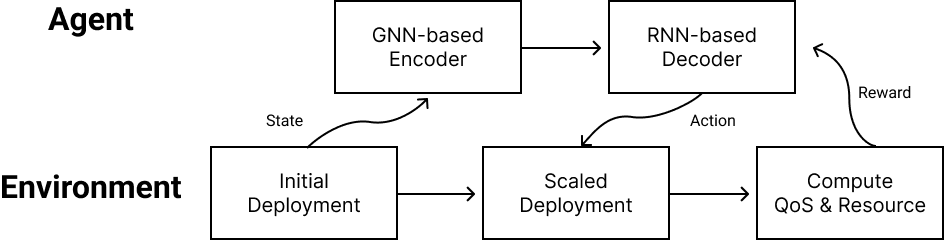}
    \caption{Overview of the RL pipeline for VNF scaling: 1) The initial deployment is observed as the state from the environment. 2) Then, the RL agent updates the deployment by scaling action. 3) Finally, the reward is given, and one episode ends.}
    \label{fig:scaling}
\end{figure}

In the RL formulation, the RL agent observes the VNF deployment and a set of SFC requests as the initial state and produces the scaling actions for all the nodes and VNF types from the policy network. Then, after scaling from the initial VNF deployment, the reward is computed \footnote{The traversing path of the requests is generated by GNN-based SFC model \cite{SFCmodel2020Heo}.}, which is defined as follows:
\begin{equation}\label{eqn:reward_func}
    R = -\frac{1}{N_{req}}\sum_k\frac{\delta^k}{\delta_{SLA}^k} - \alpha \sum_{vnf} N_{vnf},
\end{equation}
where $N_{req}$ and $N_{vnf}$ are the number of SFC requests and the instances of VNF type $vnf \in \{Firewall, NAT, \cdots \}$. $\delta^k$ and $\delta^k_{SLA}$ are the traversing delay and SLA of request $k$, and $\alpha$ is the coefficient of penalty for restriction of resource utilization. The RL scenario is illustrated in Fig. \ref{fig:scaling}.

\section{Proposed Method}

In this paper, we propose to enhance the neural network model and adapt new learning techniques. First, we enhance the previous VNF scaling model \cite{namjin2022ppodepl} so that the model can be adaptable to more general deployment settings which includes nodes which cannot have any VNF instances deployed on them. 
Then, we apply new RL algorithms like PPG to train the model more efficiently.

\subsection{Model Architecture} \label{sec:policynet}

\begin{figure}[h]
    \centering
    \begin{subfigure}[b] {\textwidth}
    \includegraphics[width=\textwidth]{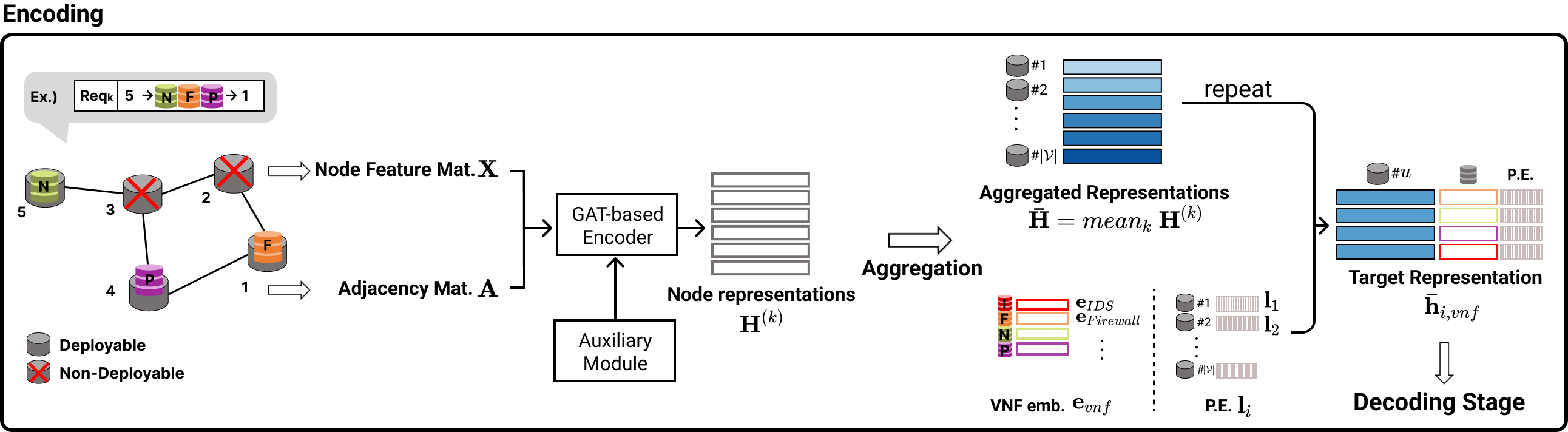} 
    \caption{Encoding}
    \end{subfigure}
    \hfill
    \begin{subfigure} [b] {\textwidth}
    \includegraphics[width=\textwidth]{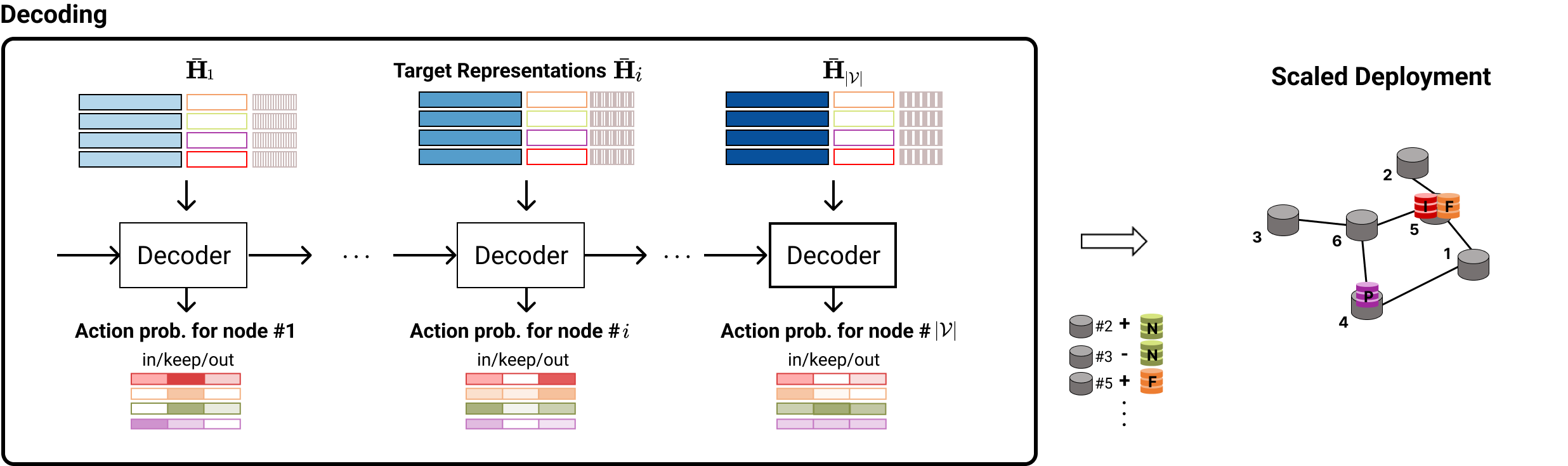}
    \caption{Decoding (Example of Firewall as target VNF type)}
    \end{subfigure}
    \caption{The proposed model architecture and the pipeline. (a) Node representations $\{\mathbf{H}^{(k)}\}$ are obtained from all the pairs of an SFC request and the VNF deployment, and the target representation $\mathbf{\bar h}_{i,vnf}$ for the target VNF type $vnf$ on target node $v_i$ is obtained by averaging the node representations $\{\mathbf{H}^{(k)}\}$ and concatenating with target VNF embeddings $\mathbf{e}_{vnf}$ as well as a positional encoding $\mathbf{l}_i$ (``P.E.''). (b) The decoder reads the target representation $\mathbf{\bar h}_{i,vnf}$ to make a decision for scaling action of the target at the decoding step $i$. Over every decoding step, the decoder generates actions for the target VNF type, and the iteration stops after the number of target nodes. Scaling for all the VNF types can be performed in parallel.}
    \label{fig:model}
\end{figure}

As in \cite{namjin2022ppodepl}, the encoder-decoder based model reads the network information and produces the action probabilities for all the nodes and VNF types. The pipeline of the model architecture is presented in Fig. \ref{fig:model}.

The encoder computes the node representations $\mathbf{H}^{(k)}$ by forwarding the pair of the adjacency matrix $\mathbf{A}$ and node feature matrix $\mathbf{X}^{(k)}$ for each SFC request, where $\mathbf{X}^{(k)}=[\mathbf{x}_1^{(k)}, \cdots, \mathbf{x}_{|\mathcal{V}|}^{(k)}]$ is the node feature matrix for SFC request $k$. Specifically, node $v_i$'s feature is defined by 
\begin{equation}
\label{eqn:node_feature}
\mathbf{x}^{(k)}_i=[src^{(k)}(i), N^{(k)}_{i, Firewall}, N^{(k)}_{i,NAT}, \cdots, dst^{(k)}(i)],
\end{equation}
where $src^{(k)}(i)$ and $dst^{(k)}(i)$ indicate whether node $v_i$ is the ingress or egress of request $k$. $N^{(k)}_{i,Firewall}, N^{(k)}_{i,NAT}, \cdots$ are the instance numbers of VNF type $vnf \in \{Firewall, NAT, \cdots \}$ deployed on node $v_i$ if needed for request $k$, otherwise set to 0. After encoding all the SFC requests, the encoder outputs the sets of node representations $\{\mathbf{H}^{(k)}\}$. 
Then, $\{\mathbf{H}^{(k)}\}$ are averaged into a single representation $\mathbf{\bar H}$, which is broadcasted and concatenated with each VNF embeddings $\mathbf{e}_{vnf} \in \{\mathbf{e}_{Firewall}, \mathbf{e}_{NAT}, \cdots  \}$ to make target VNF representations $\mathbf{\bar h}_{i,vnf}$ at the target nodes.


The RNN-based decoder consisting of GRU and MLP layers reads the target representation $\mathbf{\bar h}_{i,vnf}$ at each decoding step $i$, then outputs the action probabilities of the target VNF type $vnf$ on target node $v_i$. The actions include three types of scaling actions for the target: `scale-in', `scale-keep', and `scale-out'. Over every decoding step, the decoder generates actions for the target VNF type, and the iteration stops after the number of target nodes. Scaling for all the VNF types can be performed in parallel, with this single process.


\subsubsection{GAT-based Encoder}

We apply the graph attention network (GAT) for the encoder instead of GGNN \cite{SFCmodel2020Heo}.  
GAT has several training advantages compared to GGNN for VNF deployment. 
First, GAT could effectively propagate the node information from the previous layers through the attention mechanism, while GGNN relies on RNN-like updates. 
Next, GAT updates the node representation with different importance of the neighbor nodes as the attention score, which is helpful to train the model for requests with multiple hops between ingress and egress. Due to the fact that the VNF deployment is highly related to the paths of the requests, it is essential to train the importance of neighbors on the paths.


The input of the GAT encoder is the adjacency matrix $\mathbf{A}$ and the input node feature matrix $\mathbf{X}$ given the input pair of the VNF deployment and an SFC request. The encoder produces node representation matrix $\mathbf{H} = \left[\mathbf{h}_1, ... , \mathbf{h}_{|\mathcal V|}\right]$.
The input node features and edge weights defined as link latency are first forwarded through a shared linear transformation, parameterized by weight matrices $\mathbf{W}, \mathbf{W}_e$, where $\mathbf{W}, \mathbf{W}_e$ are weight matrices for node features and edge latency. Then, self-attention is performed on the nodes, where attention scores $\alpha_{ij}$ between node $v_i$ and $v_j$ are computed only for neighbor nodes, which is defined by Eq. \ref{eqn:gat2}. In addition, we employ multi-head attention to stabilize the learning process \cite{velickovic2018gat}. Finally, the node representation $\mathbf{h}_i$ is updated as a weighted sum with the attention scores over nodes.

\subsubsection{Auxiliary Encoding Module with Node Embedding} \label{sec:nodeemb}

\begin{figure*}[h]
    \centering
    \begin{subfigure}[t] {0.4\textwidth}
    \includegraphics[width=\textwidth]{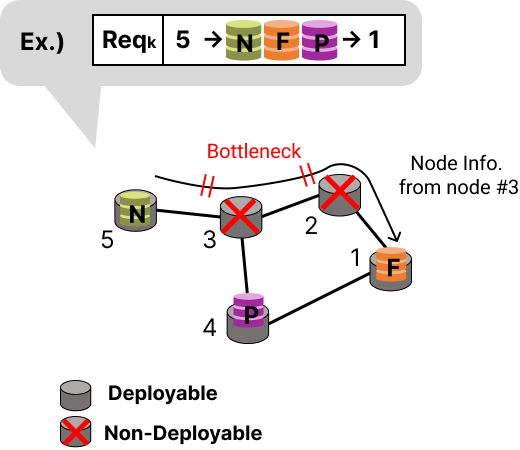} 
    \caption{Illustrative example of how non-deployable nodes could be bottlenecks, which hinder the information propagation from node\#5 to node\#1.}\label{fig:nodeemba}
    \end{subfigure}
    \hspace{0.2em}
    \begin{subfigure} [t] {0.53\textwidth}
    \includegraphics[width=\textwidth]{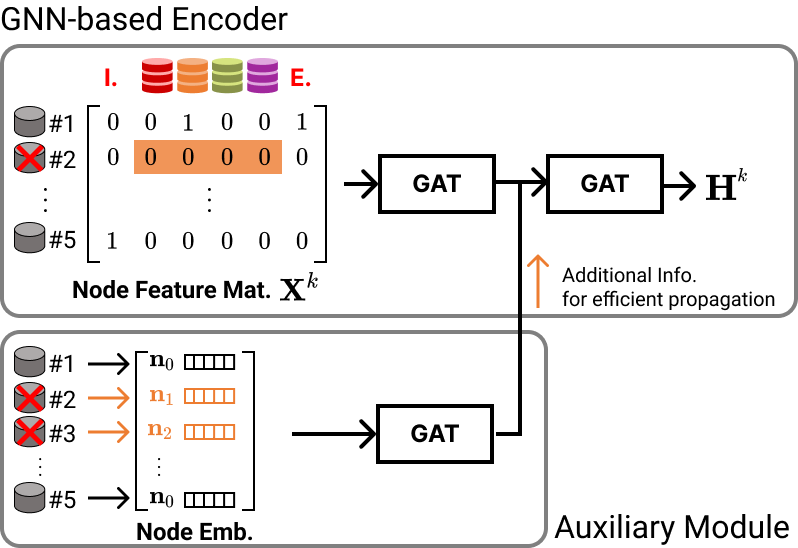}
    \caption{The auxiliary module provides the neighboring information of non-deployable nodes so that the GNN-based encoder could update node representation efficiently.}\label{fig:nodeembb}
    \end{subfigure}
    \caption{Example of the general deployment setting and architecture of the encoder.}
    \label{fig:hetero}
\end{figure*}

In the general deployment setting, the nodes in the network can be classified into 2 different types: non-resource-allocating nodes (``non-deployable'') and resource-allocating nodes (``deployable''). As defined in Eq. \ref{eqn:node_feature}, node feature $\mathbf{x}_i$ contains the number of instances for each VNF type, which are always zero-valued for the non-deployable nodes ($N^{(k)}_{i, vnf}=0$). Fig. \ref{fig:nodeemba} represents an example of node features.

Given the different node types, the model cannot update enough the representation of non-deployable nodes since the zero-valued features cannot have enough attention, which could cause a bottleneck when propagating node information through these nodes. 
To overcome the issue, we designed an auxiliary encoding module that provides additional neighbor information to the GNN-based encoder, as illustrated in Fig. \ref{fig:nodeembb}. This information is then utilized by the next module of the encoder to complement for information propagation when updating the representation of deployable nodes.

The auxiliary encoding module consists of vectorized embeddings and a GAT layer. We first initialize vectorized embeddings $\{ \mathbf{n}_1, \mathbf{n}_2, \cdots \}$ randomly and assign them to each non-deployable node, and then update them along with other parameters. As these embeddings are optimized during the training process, the model could provide the non-deployable nodes' neighboring information to mitigate the bottleneck effect. We refer to these embeddings as node embeddings (`N.E.'). In addition, we define and assign the same node embedding $\mathbf{n}_0$ to all deployable nodes, allowing the module to focus on non-deployable cases. 
Subsequently, we process these node embeddings through a GAT layer to further reflect relations between multi-hops neighbors. We implement the GAT layer of this auxiliary module with less parameters than the encoder's GAT layers to avoid over-fitting.

Finally, we concatenate the output of the auxiliary module and the output of the GNN-based encoder's first layer (refer to Fig. \ref{fig:nodeembb}) and forward to the second layer of the encoder to get a final nodes representations $\mathbf{H}^k$. 

\subsubsection{Positional Encoding}
To pass spatial information of nodes, we employ positional encoding, as in the Transformer \cite{vaswani2017transf}. Positional encoding could alleviate the decoder's burden of transferring knowledge from previous nodes to decode hidden states.
Before forwarding the aggregated representation $\mathbf{\bar H}$, the location vector $\mathbf{l}_i$ is computed with sine and cosine functions of different frequencies as in the Transformer, then concatenated with $\mathbf{\bar h}_i$. 
Finally, the decoder inputs the representation of node $v_i$ partitioned into three parts: aggregated node $v_i$'s representation $\mathbf{\bar h}_i$, target VNF embeddings $\mathbf{e}_{vnf}$, and location vector $\mathbf{l}_{i}$.

\subsection{RL for VNF Deployment} \label{sec:ppg}
 
In this section, we discuss how we optimize the policy network with RL algorithms. We follow the setting for RL formation and the perturbation scenario as in \cite{namjin2022ppodepl}. In the perturbation scenario where the VNF deployment is perturbed from the optimal deployment, the RL agent needs to remove the redundant VNFs or add necessary VNF instances to reconstruct optimal deployment by making scaling decisions for all the nodes and VNF types in the current deployment. 

In the perturbation scenario, we optimize the proposed model as the policy network. Moreover, we propose an architecture of the value network to apply the PPG algorithm, with which the policy network can leverage the shared representation of the SFC generation from the value network. The new value network architecture is described in the next section.

\subsubsection{Auxiliary Value Head for PPG}
The PPG algorithm jointly optimizes the objectives of the policy network and value function, where the policy network and value function share the parameters. In the optimization process, the objective of the value function is to minimize the error of the estimation for the state value, and the value function extracts the useful representation for the estimation of the SFC generation. The policy network can incorporate the representation from the value function to achieve a better policy. 

However, to avoid a conflict between both objectives of the policy and value network, PPG utilizes the auxiliary value head $V_{aux}(s)$ as well as the value network $V(s)$ to estimate the state value. The value head shares the parameters with the policy network, while the value network is implemented with the separated parameters. Thus, we design these two value functions and then apply the PPG to update the policy network.

\begin{figure}[h]
    \centering
    \includegraphics[width=0.7\textwidth]{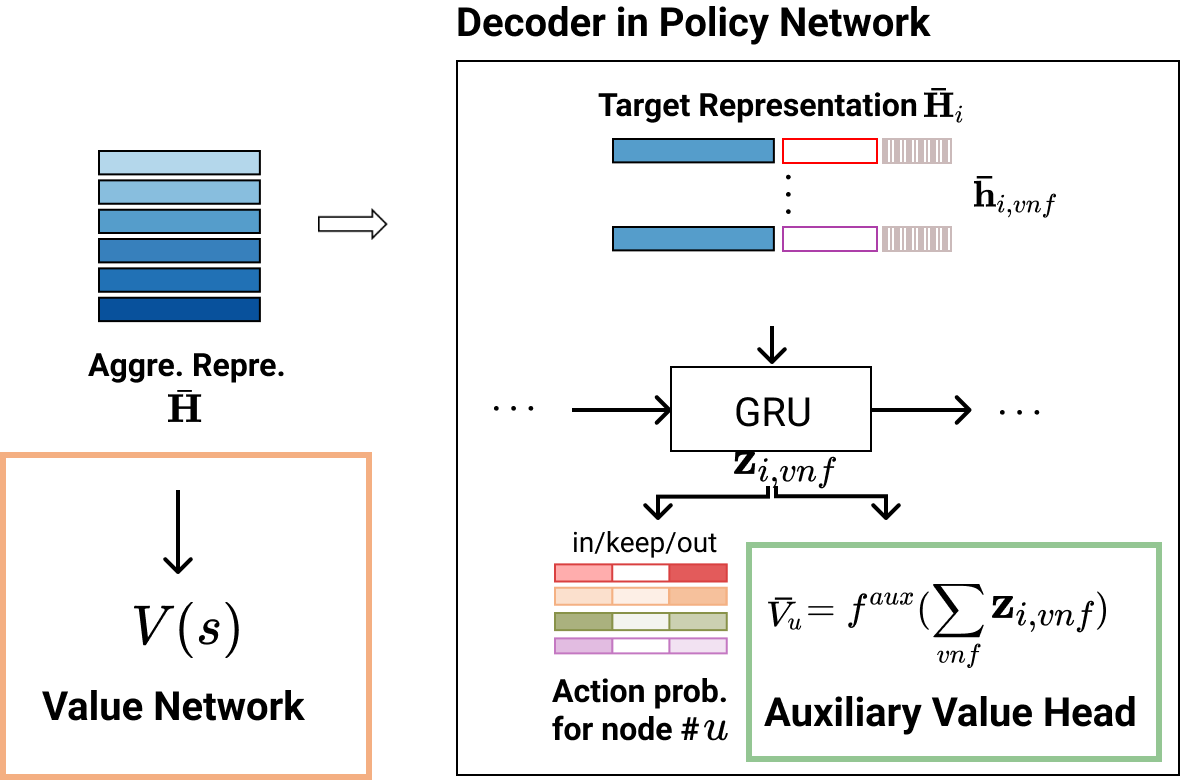}
    \caption{Architectures of the value network and the auxiliary value head: PPO optimizes the policy and the value networks (orange box). In addition, PPG optimizes the auxiliary value head (green box) sharing parameters with the policy network.}
    \label{fig:ppg}
\end{figure}

Fig. \ref{fig:ppg} presents the architectures of the value network and the auxiliary value head. 
The value network is implemented as a 2-layer MLP forwarding the mean of the aggregated representations $\mathbf{\bar H}$ to produce the state value $V(s_t)$.
For the auxiliary value head, we design the architecture to reflect the action's state value for the node at each decoding step. The value head is implemented as a 2-layer MLP which is connected to the GRU layer of the decoder of the policy network. The value head estimates $V_i$ when the policy network produces the node $v_i$'s action probabilities. Finally, the state value $V_{aux}(s_t)$ is computed by averaging $\bar V_i$ over nodes. This can be formulated by
\begin{align}
    &V_{aux}(s) = \frac{1}{|\mathcal{V}|} \sum_{v_i \in \mathcal{V}} V_i,
\end{align}
where $V_i=f^{aux}(\sum_{vnf} \mathbf{z}_{i, vnf})$, $\mathbf{z}_{i,vnf}$ is the output of the decoder's GRU layer from target VNF representation $\mathbf{\bar h}_{i, vnf}$, and $f^{aux}$ is the MLP layer of the auxiliary value head. 

We apply the PPG algorithms with the value network and the auxiliary value head. After one episode, the agent gets reward $R$ from the updated deployment, and then it stores the tuple $(R, s_t, \pi_{\theta}(a_t|s_t))$ in the buffer. Then, for $N_{PPO}$ episodes, the agent updates the parameter of the policy network to optimize the surrogate objective defined in Eq. \ref{eqn:obj_ppo}. At the same time, the RL agent updates the parameter of the value network to minimize the mean squared error (MSE) between the return and the estimated state value. Lastly, for $N_{PPG}$ episodes, the parameter of the auxiliary value head is updated to minimize the auxiliary loss, defined in Eq. \ref{eqn:aux_loss}, which jointly updates the parameter of the policy network.

\section{Experiments}

\subsection{Dataset and Configuration}
\begin{figure*}[h]
    \centering
    \begin{subfigure}[b]{0.43\textwidth}
        \centering
        \includegraphics[width=\textwidth]{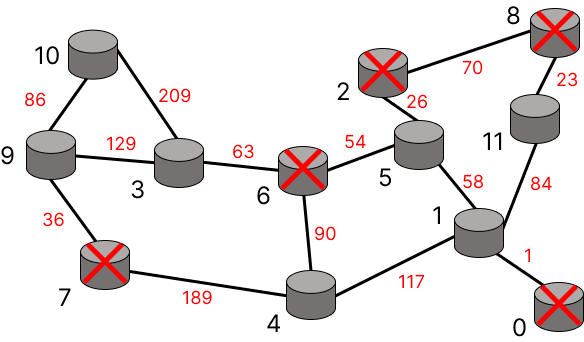}
        \caption{Internet2} \label{fig:inet2}
    \end{subfigure}
    \hspace{1em}
    \begin{subfigure}[b]{0.52\textwidth}
        \centering
        \includegraphics[width=\textwidth]{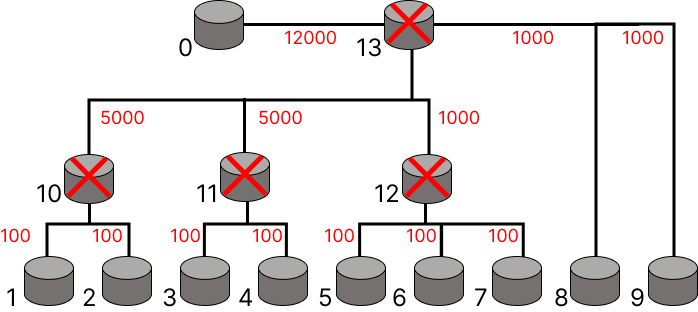}
        \caption{MEC} 
    \end{subfigure}
    \caption{Internet2 and MEC networks in the experiment settings. The nodes with `X' are non-deployable nodes like switches.} 
    \label{fig:networks}
\end{figure*}

In this section, we describe the experiment settings and configurations to train the proposed model. 
The dataset was created from two networks: Internet2 \cite{internet2tm} and Mobile Edge Computing (MEC), represented in Fig. \ref{fig:networks}. In the Internet2 network, the topology consists of 12 nodes with 15 links. SFC requests are created by normalizing Abilene traffic matrices proposed by \cite{lange2017multi}. In the MEC network, the topology consists of 14 nodes with 13 links. We followed \cite{lange2017multi} for other configurations, and specifically for the generation of SFC requests, we set the SLA as 95\% of the latency computed from the SFC path generated by ILP \cite{bari2015ilp}. 
The SFC requests contain the ingress and egress nodes, and the SFC path includes 3 or 4 services out of 5 VNF types. 

First, we generated the ILP-based deployment to determine the optimal number and location of VNF instances for the set of active requests at each interval. 
Then, we created a VNF deployment dataset of which each entry contains an ILP-based deployment and a set of SFC requests. The dataset was then divided into training, validation, and testing sets with the ratio of 8:1:1. Lastly, we implemented a simulation environment for the same topology with each network and calculated the latency of the requests in the same manner as \cite{SFCmodel2020Heo}.

Then, we trained the models on perturbed deployments by perturbing the ILP-based deployment. To perturb the deployment, an integer noise -1, 0 or +1 is added on nodes and VNF types, and these deployments are set as the initial states. Furthermore, we made random and zero deployments to evaluate the generalizability of the models. We set the number of VNF instances as 0 or 1 for the random deployment and only 0 for the zero deployment.

For the details of hyper-parameters, we mainly followed the setting of \cite{SFCmodel2020Heo} to pre-train the encoder and the configurations shown in Table \ref{tab:config} to train the decoder. The decoder architecture is implemented with layer normalization, 1-layer GRU with 64 hidden units. We use three linear layers with the same number of hidden units in the decoder for the value function. As the activation function, ReLU (Rectified Linear Unit) is used.
On each epoch, we measured the average reward in the validation set, and after the training, we selected the final model with the best reward in the validation set. 
\begin{table}[h]
\caption{Hyper-Parameters for Training.} 
\centering
\footnotesize
\begin{tabular}{l|c}
\hline
\textbf{Parameter}  & \textbf{Value} \\ \hline \hline
Learning rate (LR)\footnote{For the value network, PPO uses same LR with the policy network and PPG set the LR: 1.5e-4}   &   3e-4   \\ \hline
Decoder dim. (GRU)    &   32   \\ \hline
Decoder dim. (decoder-layer1) &   32   \\ \hline
Decoder dim. (decoder-layer2) &   32   \\ \hline
Decoder dim. (VNF embedding) &   5   \\ \hline
Decoder dim. (Positional Encoding) &   4   \\ \hline \hline
Discount factor ($\gamma$)    &   0.995 \\ \hline
PPO value network dim. (layer1)    &   128   \\ \hline 
PPO value network dim. (layer2)    &   64   \\ \hline
PPO aux. value head  dim. (layer1)    &   32   \\ \hline
PPO aux. value head  dim. (layer2)    &   32   \\ \hline
PPO epsilon ($\epsilon$)    &   0.2   \\ \hline
PPO epoch ($K$)   &   4   \\ \hline
PPO minibatch size ($M$)   &   4   \\ \hline
PPO interval \& PPG interval (policy phase) ($N_{PPO}$)    &   16   \\ \hline
PPG interval (auxiliary phase) ($N_{PPG}$)    &   64   \\ \hline
PPG hyper-parameter ($\beta_{clone}$)    &   1   \\ \hline
\end{tabular}\label{tab:config}
\end{table}

\subsection{Quantitative Results}
We evaluate the performance of the method in both networks: Internet2 and MEC. 

\subsubsection*{Internet2 Network:}
Table \ref{tab:result_inet} shows an ablation study on Internet2 where we trained the models with PPO and tested them.
For the metric, we measured the average number of VNF instances (`Avg. \#VNF'), the average of delay time, and the average SLA violation ratio (`Avg. SLAV'), as well as a reward (Eq. \ref{eqn:reward_func}). Furthermore, we train the agent with different coefficient value ($\alpha$) of instance penalty in Eq. \ref{eqn:reward_func} to show the effect of restriction for the level of the resource utilization, which controls the trade-off between the resource utilization and the QoS. We trained the agent with different seeds three times and reported the averages. 

\begin{table}[h]
\centering
\scriptsize
\begin{tabular}{l|c||rrrr} \hline
Models (SFC) & $\alpha$ &Reward &Avg. \#VNF &Avg. Delay &Avg. SLAV \\\hline\hline
ILP (GGNN) & - &-0.922 &12.86 &575.57 &0.197 \\
ILP (GAT) & - &-0.881 &12.86 &545.42 &0.167 \\
ILP (GAT+N.E.) & - &-0.876 &12.86 & 542.11 &0.162 \\
\hline
GGNN (GGNN) &0.15 &-1.796 &19.89 &1158.01 &0.279 \\
GGNN (GGNN) &0.20 &-1.877 &18.44 &1226.94 &0.260 \\
GGNN (GGNN) &0.25 &-2.345 &16.57 &1580.09 &0.354 \\
\hline
GAT(GAT) &0.15 &-1.376 &17.81 &868.16 &0.141 \\
GAT(GAT) &0.20 &-1.580 &17.58 &1018.22 &0.179 \\
GAT(GAT) &0.25 &-2.018 &16.40 &1343.65 &0.255 \\
\hline
GAT+N.E. (GAT+N.E.) &0.15 &-1.062 &17.71 &641.21 &0.086 \\
GAT+N.E. (GAT+N.E.) &0.20 &-1.314 &16.04 &836.31 &0.117 \\
GAT+N.E. (GAT+N.E.) &0.25 &-1.345 &15.75 &860.86 &0.122 \\
\hline
GAT+P.E. (GAT) &0.15 &-1.084 &16.59 &665.61 &0.093 \\
GAT+P.E. (GAT) &0.20 &-1.283 &15.89 &815.00 &0.122 \\
GAT+P.E. (GAT) &0.25 &-1.409 &15.17 &911.44 &0.135 \\
\hline
GAT+P.E.+N.E. (GAT+N.E.) &0.15 &-1.029 &17.39 &619.91 &0.081 \\
GAT+P.E.+N.E. (GAT+N.E.) &0.20 &-1.240 &15.85 &784.09 &0.108 \\
GAT+P.E.+N.E. (GAT+N.E.) &0.25 &-1.499 &14.87 &978.89 &0.139 \\
\hline
\end{tabular}
\caption{Comparison of different models in the Internet2 network. The models are used to adjust VNF deployment based on scaling in/out/keep, while the methods in the parentheses indicate how the SFC path was created for evaluation.}
\label{tab:result_inet}
\end{table}

Each row indicates the method to make VNF deployments and the method to generate SFC paths. For example, the row 
``\textbf{ILP} (GAT)" shows the results when VNFs were deployed by the ILP method and the deployments were evaluated on the path generated by GAT for SFC. 
For the RL models, the same encoders were used for the policy networks and for the SFC path generation, though positional encoding was not used for SFC. 

From the table, we can see that the GAT-based models (`GAT') outperform the baseline models (i.e., GGNN-based models). For example, `GAT (GAT)' decreases the SLA violation rates from 0.279 (by `GGNN(GGNN)') to 0.141 when the coefficient $\alpha$ for the instance penalty is 0.15. Furthermore, it improved the level of QoS and decreased the delay time with less number of instances.
We believe that GAT models can be trained more efficiently, while GGNN-based models are hard to be optimized on our deployment settings, including many non-deployable nodes. 

Moreover, `GAT+N.E.' models outperform GGNN as well as GAT models. They decrease the SLA violation rate, the number of instances and the delay time. This means that the node embedding-based approach, described in Section \ref{sec:nodeemb} can get better representations separating the VNF deployable and non-deployable nodes. 
In addition, the positional encoding (`GAT+P.E.') shows an even further improvement on SLAV compared to the corresponding methods. It implies that the positional encoding provides more distinguishable information for the policy network to find a better policy for deployment.
Lastly, we trained GAT models with both positional encoding and node embedding (GAT+P.E.+N.E.). This method increases the reward and SLAV, while keeping the same level of the other metrics (`Avg. \#VNF' and `Avg. Delay'), compared to the GAT-based models using only the positional encoding.  


\subsubsection*{MEC network:}
We experimented in the MEC network, where 5 nodes out of a total of 12 nodes are non-deployable. We train the models with PPG, and compare GAT-based models to ILP approaches, which is summarized in Table \ref{tab:result_mec}. The reward is computed using the SFC model, which employs the same architecture as the policy network.
In the table, we can see that Positional Encoding (`GAT+P.E.') significantly decreases the SLA violation rate, the number of instances as well as the delay time. 
For the node embedding-based approach, `GAT+P.E.+N.E.' has similar performance as `GAT+P.E.'. It means node embedding does not have additional information compared to positional embedding. Actually, as shown in Fig. \ref{fig:networks}, MEC has a regular structure separating deployable and non-deployable nodes, which might be a possible explanation of why `N.E.' does not additionally increase the performance on top of `GAT+P.E.'

\begin{table}[h] 
\centering
\scriptsize
\begin{tabular}{l|c||rrrr} \hline
Models (SFC) & $\alpha$ &Reward &Avg. \#VNF &Avg. Delay &Avg. SLAV \\\hline\hline
ILP (GAT) & 0 &-0.702 &13.00 &29097.86 &0.046 \\ 
ILP (GAT+N.E.) & - &-0.707 &13.00 &29282.68 &0.045 \\
\hline
GAT (GAT) &0.15 &-1.658 &20.33 &74682.66 &0.121 \\
GAT (GAT) &0.20 &-1.867 &18.75 &87157.00 &0.146 \\
GAT (GAT) &0.25 &-2.136 &18.75 &97987.34 &0.174 \\
\hline
GAT+P.E. (GAT) &0.15 &-0.887 &16.84 &37044.12 &0.052 \\
GAT+P.E. (GAT) &0.20 &-0.956 &16.25 &41100.82 &0.060 \\
GAT+P.E. (GAT) &0.25 &-1.316 &15.41 &58091.53 &0.096 \\
\hline
GAT+P.E.+N.E. (GAT+N.E.) &0.15 &-0.915 &16.55 &38930.56 &0.055 \\
GAT+P.E.+N.E. (GAT+N.E.) &0.20 &-1.119 &16.29 &48965.09 &0.075 \\
GAT+P.E.+N.E. (GAT+N.E.) &0.25 &-1.350 &15.95 &60878.52 &0.099 \\
\hline
\end{tabular}
\caption{Comparison of different models in the MEC network. The positional encoding helps the training process more effective, while the node embedding does not additionally increase the performance on top of `GAT+P.E.'.}
\label{tab:result_mec}
\end{table}

For the GGNN-based approach, it does not optimize the reward effectively, because links of the MEC network contain high latency\footnote{The edge attribute is set as the reciprocal of latency.} and high variance, which hinders updating node representation in the model. Especially, GGNN uses a weighted sum of its node representation with the edge attribute when updating the node representation, which causes it to underestimate of the information from nodes connected with high latency. 
On the other hand, the GAT-based approach employs the attention mechanism to compute the different importance of its neighbors and the edges so that it could mitigate the underestimation problem.

\subsubsection{Comparison of Various RL algorithms}

Since we propose to use the PPG algorithms, we compare them to other RL algorithms in the Internet2 network. In the experiment, we trained the GAT-based models with positional encoding with the various RL algorithms. As a result, the PPG algorithms can obtain better rewards, compared to DQN and PPO. PPG decreases the SLA violation rate and the delay time with a similar number of VNFs as presented in Table \ref{tab:result_ppg}.
\begin{table}[h]\centering
\footnotesize
\begin{tabular}{l|c||rrrr} \hline
Method & $\alpha$ &Reward & Avg. \#VNF &Avg. Delay &Avg. SLAV \\ \hline \hline
DQN&0.15 &-1.317 &32.00 &718.56 &0.112 \\
DQN&0.20 &-1.311 &25.00 &765.22 &0.174 \\
DQN&0.25 &-2.701 &21.90 &1793.06 &0.285 \\ \hline
PPO&0.15 &-1.084 &16.59 &665.62 &0.093 \\
PPO&0.20 &-1.283 &15.89 &815.00 &0.122 \\
PPO&0.25 &-1.409 &15.17 &911.44 &0.135 \\ \hline
PPG&0.15 &-1.053 &16.30 &644.77 &0.091 \\
PPG&0.20 &-1.164 &16.19 &726.24 &0.101 \\
PPG&0.25 &-1.353 &15.51 &868.61 &0.130 \\ \hline
\end{tabular}
\caption{Comparison of the various RL algorithms. PPG can optimize the reward more effectively compared to DQN and PPO.}
\label{tab:result_ppg}
\end{table}

\begin{figure}[t]
    \centering
    \includegraphics[width=0.65\textwidth]{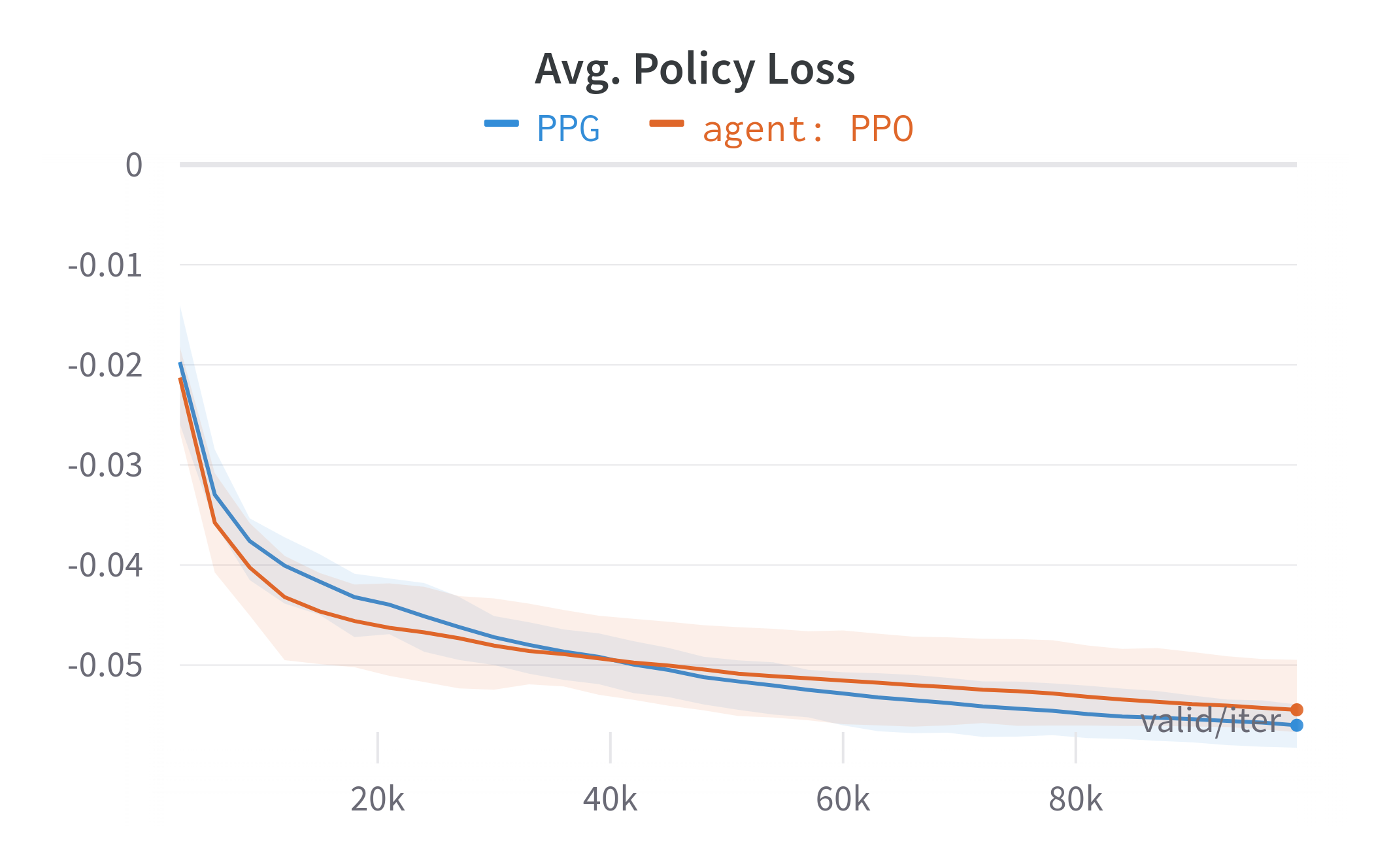}
    \caption{Comparison of policy loss ($\alpha=0.2$) in training process. The loss of PPG decreases faster than PPO after around 40k iterations.} 
    \label{fig:polloss}
\end{figure}

Moreover, as shown in Fig. \ref{fig:polloss}, PPG can optimize the average loss of the policy network more effectively than PPO. We believe that the auxiliary value head helps the policy network converge more efficiently, even though parameter sharing might make the training slow at the beginning of the training process. 

%

\subsubsection{Analysis with Random and Zero Deployments}

To analyze further the performance of the trained models, we trained the GAT-based model ($\alpha=0.2$) optimized with PPG as before, and tested the trained model on different settings, 
where initial deployment is random or no VNFs are deployed (random deployment or zero deployment). The experiment results are presented in Table \ref{tab:result_robust}, where 
``Random" and ``Zero" initial deployments are compared to the ILP-perturbed initial deployment. 
As the result, the performance on both initial deployments shows the same level of QoS and resource utilization as the ILP-perturbed case. It demonstrates that our approach can work on any sub-optimal initial deployment. Furthermore, our method can deploy VNFs without the ILP-based initial deployment. 
\begin{table}[t]\centering
    \footnotesize
    \begin{tabular}{l||rrrr} \hline
    Initial Deployment &Reward & Avg. \#VNF &Avg. Delay &Avg. SLAV \\ 
    \hline \hline
    ILP-perturbed &-1.029 &17.03 &622.59 &0.087 \\
    Random & -1.060& 16.95 & 646.12 & 0.090 \\
    Zero & -1.060& 16.94 & 646.00& 0.090\\ \hline
    \end{tabular}
    \caption{Evaluations from different initial deployments in testing for the trained GAT model with $\alpha=0.2$.}
    \label{tab:result_robust}
\end{table}

\subsubsection{Execution Time}
To show how fast our method works, we measured the average execution times to make decisions for VNF deployments and compared the proposed GAT-based model to ILP. 
The execution time includes the VNF deployment actions by the agent as well as SFC path generation by the SFC model \cite{SFCmodel2020Heo}.
Each approach makes 10 deployments on a machine with 24 x 12-Core AMD Ryzen 9 3900X CPUs. Table \ref{tab:computation} presents the average results of each approach. Our method can process VNF deployment about 30 times faster than ILP. 
\begin{table}[t]\centering
    \footnotesize
    \begin{tabular}{l||rr} \hline
     Method & Avg. Time (sec) \\
     \hline \hline 
     ILP & 14.06 \\
     GAT & 0.431  \\
     \hline
    \end{tabular}
    \caption{Average of execution times of 10 VNF-deployments.}
    \label{tab:computation}
\end{table}

\subsection{Qualitative Evaluation}
In this section, we discuss the quality of the results by our approach. We analyze the VNF deployment generated from the agent and plot t-SNE \cite{lauren2009tsne} of the representation computed from the encoder of the policy networks. 

\subsubsection{Generated Deployment}
 
\begin{figure}
    \centering
    \includegraphics[width=\textwidth]{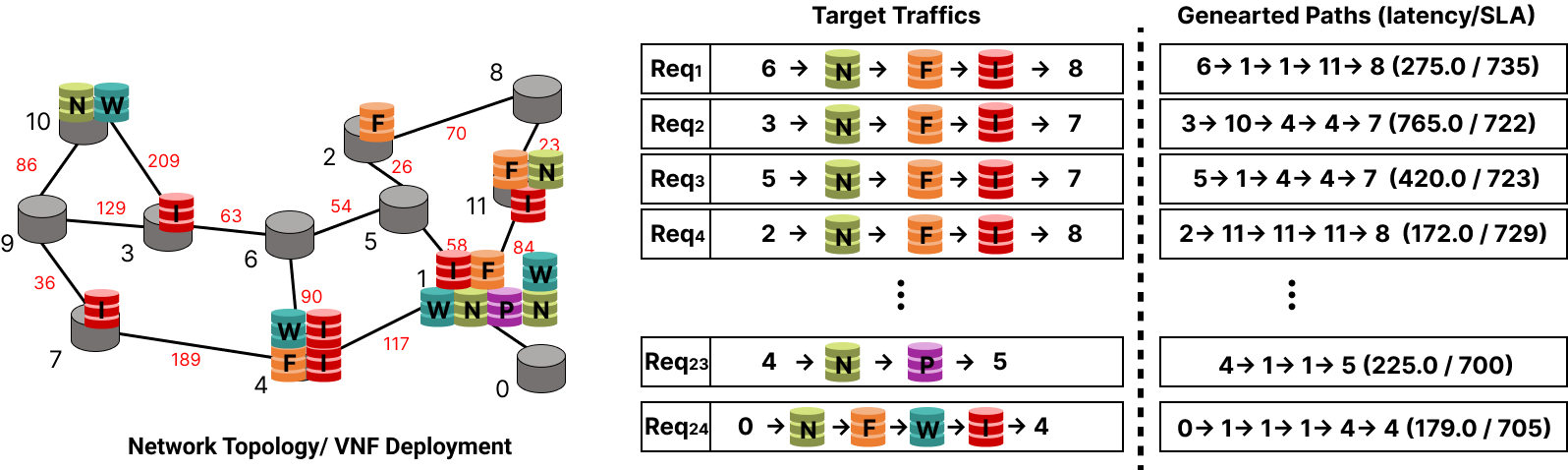}
    \caption{The VNF deployment generated from the model (Left), target SFC requests (Middle), and generated paths (Right). Given the target SFC request, VNF deployment and path are generated by the proposed method and the SFC model.}
    \label{fig:gen_depl}
\end{figure}

We plot the VNF deployment generated from the GAT-based model trained with PPG as presented in Fig. \ref{fig:gen_depl}, which shows VNF deployment generated from the SFC requests, and paths generated for the requests by the SFC model \cite{SFCmodel2020Heo}. The VNF deployment is generated with more than 20 requests, so we omitted the other requests in the figure for simplicity.

The most VNF instances are deployed in the middle of the shortest paths for the traffics. For example, the first traffic (`$\mathbf{Req}_1$') has node 6 and node 8 as the ingress and egress, and is required to pass through NAT (`N'), Firewall (`F'), and IDS (`I'). In the network, the traffic goes through NAT at node 1, Firewall at node 11, and IDS at node 8. This traffic takes 275 ms while SLA for the SFC path is 735 ms. That is, all the VNFs of `$\mathbf{Req}_1$' are deployed on the shortest path of its ingress and its egress. 

In addition, the generated deployment needs to meet the QoS with the optimized amount of resources. As shown in Fig. \ref{fig:gen_depl}, VNF instances are deployed on the intersection of the shortest paths for the traffics. For example, the VNF instances at node 1 process more than four requests, like NAT for `$\mathbf{Req}_1$', `$\mathbf{Req}_2$', `$\mathbf{Req}_{23}$' and `$\mathbf{Req}_{24}$'. That is, our approach can deploy the instances at the shared nodes of the paths for many requests, which can reduce the number of VNF instances.

\subsubsection{Node Representations by t-SNE}

\begin{figure*}[h]
    \centering
    \begin{subfigure}[b]{0.4\textwidth}
        \centering
        \includegraphics[width=\textwidth]{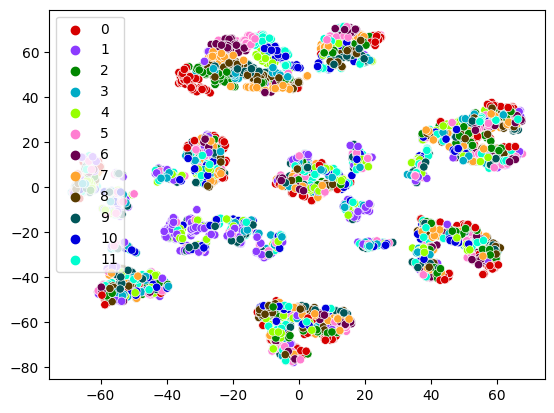}
        \caption{GGNN}\label{fig:tsne_ggnn}
    \end{subfigure}
    \hspace{1em}
    \begin{subfigure}[b]{0.4\textwidth}
        \centering
        \includegraphics[width=\textwidth]{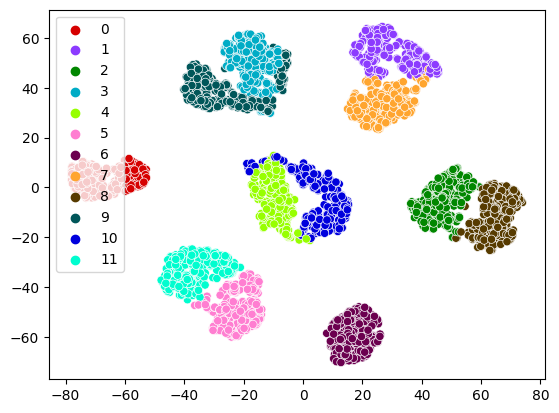}
        \caption{GAT}
        \label{fig:tsne_gat}
    \end{subfigure}
    \caption{Node representations by t-SNE. The GAT-based model gets more disentangled representations, compared to the GGNN-based model.} 
    \label{fig:tsne}
\end{figure*}

In the section, to analyze how the network information is represented by the encoder, we plot the 2D representation of node embeddings by t-SNE \cite{lauren2009tsne} as in Fig. \ref{fig:tsne}, which shows the representations from the GGNN-based encoder and our proposed GAT-based encoder. Each dot represents one node of network, and the nodes are located in the Internet2 network as shown in Fig. \ref{fig:inet2}. 

Even though the GGNN-based encoder has multiple clusters, the nodes within each cluster are not distinguishable. However, the GAT-based encoder's representations are disentangled so that each cluster contains only one or two nodes. Furthermore, the representations of neighbors in the network are close to each other in the plot. For example, clusters for node 8 (`brown') and node 2 (`green') are close to each other, which reflects its neighborhood with the same node type as presented in Fig. \ref{fig:inet2}. This demonstrates the representation by GAT-based model reflects the network topology effectively.


\section{Conclusion}
In this paper, we proposed enhanced models which can be adapted to more general network settings. We proposed an improved GNN architecture and a few techniques to obtain a better node representation for the VNF deployment task. Furthermore, we optimized the model with PPG, a variant policy gradient-based algorithm. In the experiment, we evaluated the proposed method in various scenarios, achieving a better QoS with minimum resource utilization compared to the previous methods. Finally, we analyzed the generated VNF deployment and compare the node representations of our model to the baseline. 

\section*{Acknowledgement}
This research was supported by Basic Science Research Program through the National Research Foundation of Korea funded by the Ministry of Education (NRF-2022R1A2C1012633), and by Institute for Information \& communications Technology Promotion (IITP) grant funded by the Korea government(MSIT) (No. 2018-0-00749, Development of virtual network management technology based on artificial intelligence).

\bibliographystyle{elsarticle-num} 
\bibliography{refs}

\end{document}